\documentclass[10pt,a4paper,english]{elsart}
\usepackage[T1]{fontenc}
\usepackage[latin1]{inputenc}
\usepackage{babel}
\usepackage{graphics}

\makeatletter

\providecommand{\LyX}{L\kern-.1667em\lower.25em\hbox{Y}\kern-.125emX\@}

 \usepackage{graphicx,natbib,theorem}
\newcommand{\maketitle}{\end{frontmatter}}

\makeatother
\begin{document}
\begin{frontmatter}

\title{Random Matrix Theory in Lattice Statistical Mechanics }

\author{J.-Ch. Anglès d'Auriac, J.-M. Maillard}

\address{CRTBT-CNRS, 25 av. des Martyrs, 38042 Grenoble FRANCE}

\address{LPTHE Tour 16, 4 Place Jussieu, F--75252 Paris Cedex 05, FRANCE}

\begin{abstract}
In this short note we collect together known results on the use of
Random Matrix Theory in lattice statistical mechanics. The purpose
here is two fold. Firstly the RMT analysis provides an intrinsic characterization
of integrability, and secondly it appears to be an effective tool
to find new integrable models. Various examples from quantum and classical
statistical mechanics are presented.
\end{abstract}
\maketitle

\section{Introduction}

The Random Matrix Theory (RMT) was introduced in the early fifties
by E.P. Wigner Ref.\cite{Wigner50} to study heavy nucleus. The key
idea was to replace Complexity by Randomness, arguing that the Hamiltonian
of a real heavy nucleus is so complicated that its full determination
is out of reach. Instead of trying to include all the physical ingredients,
one considers, in the RMT analysis, that the resulting operator can
be seen as the representative of a suitable statistical ensemble.
Of course this is only an approximation but the RMT analysis turned
out to give very good results in describing many situations in nuclear
physics. 

After this pioneering work, the RMT analysis has been applied to many
fields of physics, and also of pure mathematics. Schematically one
can draw a crude analogy with the large number law: the distribution
of the sum of a large number of independent random variables is, under
some restrictive conditions, a Gaussian; in the same way the spectrum
of a sufficiently {}``generic'' Hamiltonian is well approximated
by the {}``average'' spectrum of operator statistical ensemble.
A general presentation of the Random Matrix Theory can be found in
Refs.\cite{wignerbook,metha}.

This statistical ensemble depends on the symmetries of the physical
system under consideration. Actually this RMT analysis does not apply
to a single Hamiltonian, but to a family of Hamiltonians, i.e. an
Hamiltonian \emph{depending on some parameters}. Four statistical
ensembles are sufficient to describe the main situations one can encounter.
If the family of the operator can be expressed in a basis \emph{independent
of the parameter}s where all the entries of a symmetric matrix are
real, then the probability distribution should be invariant under
any orthogonal transformation. If one also requires independence of
the entries one is led to the so-called Gaussian Orthogonal Ensemble
(GOE), which is the set of symmetric matrices with entries drawn from
a centered, and normalized, Gaussian distribution (except diagonal
entries for which the root-mean square is two). A family of Hamiltonian
is time-reversal invariant if there exits an operator \( T \) such
that \( Te^{iH(\{\lambda \})t}T=ae^{-iH(\{\lambda \})t} \). This
condition is fulfilled iff there exists a unitary operator \( K \)
such that \( H(\{\lambda \})K=K\overline{H(\{\lambda \})} \), where
\( K \) can be either symmetric or antisymmetric and \( \overline{H} \)
denotes the conjugate. Note that any symmetric and unitary operator
\( K \) can be written as the product of a unitary operator \( U \)
and its transpose, namely \( K=U\widetilde{U} \), and thus one can
perform a change of basis bringing the hermitian Hamiltonian \( H \)
into a symmetric hermitian, and \emph{thus real}, matrix: \( U^{-1}HU \).
One sees that if an operator is time reversal invariant, and if \( K \)
is symmetric, then the GOE will apply. By contrast if \( K \) is
antisymmetric, the GOE will not apply. Instead the so-called Gaussian
Symplectic Ensemble (GSE) will apply. This is the ensemble of quaternion
hermitian matrices. In the case where the family of the operator is
not time reversal invariant, then the Gaussian Unitary Ensemble (GUE)
will apply. The probability distribution is then invariant under any
unitary transformation. This is the ensemble of hermitian matrices
with both real and imaginary parts of each entries being independent
and drawn from a Gaussian distribution. The fourth case is precisely
the very peculiar case of \emph{integrable} models. In this case there
exists a basis \emph{independent of the parameters} in which the Hamiltonian
is diagonal since there are as many commuting operators as the size
of the Hilbert space. The ensemble to introduce here is simply the
Random Diagonal matrix Ensemble (RDE), i.e. diagonal matrices with
random independent diagonal entries. 

In the next section we sketch how to apply these ideas to quantum
and classical lattice statistical mechanics. In the last section we
illustrate the RMT with various models of lattice statistical mechanics.

\section{Application to lattice statistical mechanics}

When applied to quantum statistical mechanics, it is clear that RMT
analysis has to be performed on the Hamiltonian itself. However for
classical statistical mechanics, it is not clear what is the operator
to be considered. Take for example the classical Ising model. The
possible values of the energy are \( E_{k}=-(N_{e}-2k)J \), where
\( N_{e} \) is the number of edges of the lattice and \( J \) is
the coupling constant: the spectrum is totally rigid and thus will
not be described properly by any of the four statistical ensembles
introduced in the previous section. It has been shown that the proper
operator to consider for classical models is the \emph{transfer matrix}
Ref.\cite{PreJMHJC}. Actually the procedures we present below always
apply either to the quantum Hamiltonian for the quantum model or to
the transfer matrix for the classical model.

Before performing the RMT analysis on a given family of Hamiltonian
one has to consider its symmetries. By symmetry we mean a linear operator
\emph{independent of the parameters} acting on the same Hilbert space
and which commutes with the family of Hamiltonian. The set of such
operators forms a group. Using the irreducible representations of
this group one can find a basis in which the Hamiltonian is block-diagonal,
each block defining a sector indexed by quantum numbers. Obviously,
states belonging to different sectors are not correlated and the analysis
has to be performed separately in each sector. These symmetries are
usually the lattice symmetries (i.e. the automorphy group of the lattice
in graph theory langage), the spin symmetries (for example the \( O(3) \)
spin and pseudo-spin symmetry of the Hubbard model Ref. \cite{key-50}),
and the color symmetry (for example the permutation of the states
in a \( q \)-state Potts model). The number of those symmetries is
a power of the number of sites of the lattice whereas the Hilbert
space size grows exponentially with this number. Consequently the
size of each block remains, after the block diagonalization, an exponential
function of the number of sites of the lattice. However in the very
special case of an integrable family of Hamiltonian, the number of
symmetries equals the Hilbert space size, and a total reduction would
lead to a completely diagonal matrix. In practice one does not know
all the symmetries and the block diagonalization is only partial,
leading to blocks well represented by RDE.

The density of states of the various models of lattice statistical
mechanics are very different of each other. Obviously no universality
can be found in the raw spectrum. Instead one can write the integrated
density of states as \( \rho (\lambda )\simeq \textrm{ regular}(\lambda )+\textrm{scale }\times \textrm{universal}(\lambda ) \),
where the regular part does depend on the model while the universal
part does not. The possible forms of the universal part are given
by the four ensembles described in the first section. The procedure
to extract this universal part is known as the \emph{unfolding} of
the spectrum. It has been described in many references, and amounts
to transforming the raw eigenvalues into unfolded eigenvalues, which
have a local density of states very close to one everywhere in the
spectrum.

Once the spectrum has been sorted according to quantum numbers and
properly unfolded, it remains to {}``compare'' it with the spectrum
of the four ensembles GOE, GUE, GSE or RDE. For a given Hamiltonian
the eigenvalues are well determined and the joint probability distribution
of the eigenvalues is simply a Dirac measure. It will never be the
joint probability distribution of the eigenvalues of the Gaussian
Ensembles which is well known to be\[
P_{\beta }(\lambda _{1},\cdots ,\lambda _{n})=C\prod _{i<j}\left| \lambda _{i}-\lambda _{j}\right| ^{\beta }\exp \left( -A\sum _{i}\lambda ^{2}_{i}\right) \]
where \( \beta =1,2,4 \) respectively for GOE, GUE and GSE is the
level repulsion. Instead, one can restore {}``probabilistic'' properties
introducing the level spacings. Sorting the unfolded eigenvalues in
ascending order, the set of the differences \( s=\lambda _{i}-\lambda _{i-1} \)
between consecutive eigenvalues does form a distribution which can
be compared to the four reference level spacings:\begin{eqnarray*}
P_{\textrm{RDE}}(s)=\exp (-s) & \qquad  & P_{\textrm{GOE}}(s)=\frac{\pi }{2}\exp (-\pi s^{2}/4)\\
P_{_{\textrm{GUE}}}(s)=\frac{2^{5}}{\pi ^{2}}\exp (-4s^{2}/\pi ) & \qquad  & P_{\textrm{GSE}}(s)=\frac{64^{3}}{9^{3}\pi ^{3}}s^{4}\exp (-64s^{2}/9\pi )
\end{eqnarray*}
Note that the above expression are only approximations of the corresponding
level spacing distribution%
\footnote{The actual distributions are in fact related to Painlevé transcendents.
}. In practice it is useful to use a parametrized distribution which
extrapolate between RDE and GOE. Using the following distribution

\begin{equation}
\label{PBrody}
P_{\beta }(s)=c(\beta +1)s^{\beta }\exp (-cs^{\beta +1})
\end{equation}
one can find the value of \( \beta  \) realizing the best fit: a
small value of \( \beta \sim  \) 0.1 will indicate an integrable
model, whereas a value of \( \beta \sim  \) 0.9 will indicate a GOE
statistic of the eigenvalues and consequently a time-reversal model.

If one want to test how close the given Hamiltonian is from the statistical
ensemble, one compute other quantities involving more than only two
consecutive eigenvalues. One of these quantity is the so-called rigidity\[
\Delta _{3}(L)=\left\langle \frac{1}{L}\min _{a,b}\int ^{\alpha +L/2}_{\alpha -L/2}\left( \rho (\lambda )-a\lambda -b\right) ^{2}\right\rangle _{\alpha }\]
here the bracket an average over all the possible position of the
{}``window'' of width \( L \). The behavior of the rigidity for
the RDE, GOE, GUE and GSE is known and is presented for comparison
in the figures of the next section.

\section{Examples}

This section is devoted to various examples. We will see that indeed
non integrable model compare extremely well with the corresponding
Gaussian ensemble, and also that the spectra of integrable models
are, in many respect, close to a set of independent numbers (RDE)
Refs\cite{poilbl,key-14,theseHend,PhysREvESpin,kagomeJC,key-20,key-24,key-33,key-50,key-56,key-9}.

\subsection{The Generalized Hubbard Chain.}

We begin with the generalized Hubbard Chain (see Ref.\cite{theseHend})
which describes a set of electrons (or any spin one-half particles)
on a chain and interacting via both a Coulomb repulsion \( U \),
a proximity interaction \( V \) and a Heisenberg coupling \( J \).
The results presented in this section originate in a long-standing
collaboration of H. Meyer with the authors (see Ref. \cite{theseHend}).
The Hamiltonian reads: \begin{equation}
\label{HHub}
H=t\sum _{i,\sigma }c_{i+1,\sigma }^{\dagger }c_{i,\sigma }+U\sum _{i}n_{i\uparrow }n_{i\downarrow }+V\sum _{i}n_{i}n_{i+1}+J\sum _{i}\overrightarrow{S}_{i}\: \overrightarrow{S}_{i+1}
\end{equation}
This model is particularly interesting since it may, or may not, be
integrable, depending on the parameters. In this context integrable
means that the eigenfunctions actually have the form proposed in the
Bethe ansatz (see Ref.\cite{gaudin}), or in its refined nested form.
The known integrable cases are summarized in the following table:

{\centering \begin{tabular}{|c|ccc|}
\hline 
&
\( U/t \)&
\( V/t \)&
\( J/t \)\\
\hline
\hline 
Hubbard&
\( \forall  \)&
0&
\( 0 \)\\
\hline 
\( t-J \) supersymetric&
\( \infty  \)&
\( \pm 1/2 \)&
\( \mp 2 \)\\
&
\( \infty  \)&
\( \pm 3/2 \)&
\( \pm 2 \)\\
t-0&
\( \infty  \)&
\( 0 \)&
\( 0 \)\\
\hline 
\( XXZ \) chain&
\( \infty  \)&
\( \forall  \)&
\( 0 \)\\
\hline
\end{tabular}\par}

In Fig. \ref{FigHubPdsDe3}, we compare the level spacing and the
rigidity in two paradigm cases Ref.\cite{theseHend}.
\begin{figure*}
{\centering \includegraphics{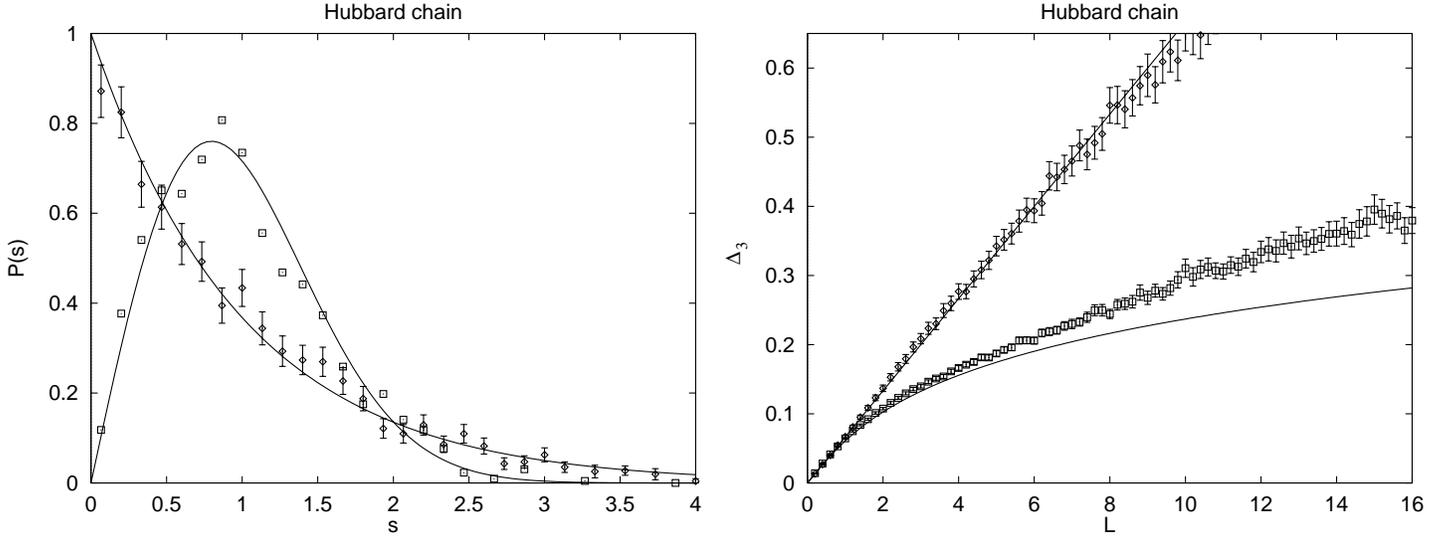} \par}

\caption{\label{FigHubPdsDe3}Generalized Hubard Chain}
\end{figure*}
One case \( t=1 \), \( U=10 \) and \( J=V=0 \) corresponds to an
integrable case, and, indeed, the level spacing \( P(s) \) and the
rigidity \( \Delta _{3} \) are in good agreement with the prediction
of independent eigenvalue (RDE), whereas the second case \( t=U=1 \),
\( V=0 \) and \( J=2 \) corresponds to a generic non integrable
case and is in good agreement with the GOE ensemble. The Hamiltonian
Eq.\ref{HHub} being real, one expects GOE rather than GUE or GSE.

To better follow how the level repulsion \( \beta  \) behaves in
the different region of the parameter space, we define a path in the
phase space, and record \( \beta  \) as we move along this path.
To be specific we simply fixed \( U=0 \) and, for different values
of \( V \), we vary \( J \). The parameter \( \beta  \) corresponds
to a best fit of the distribution Eq. \ref{PBrody}. On figure Fig.
\ref{FigHubbeta} the specific integrable points are clearly seen
as points for which the parameter \( \beta  \) drops to zero, in
excellent agreement with the previous table.
\begin{figure*}
{\centering \includegraphics{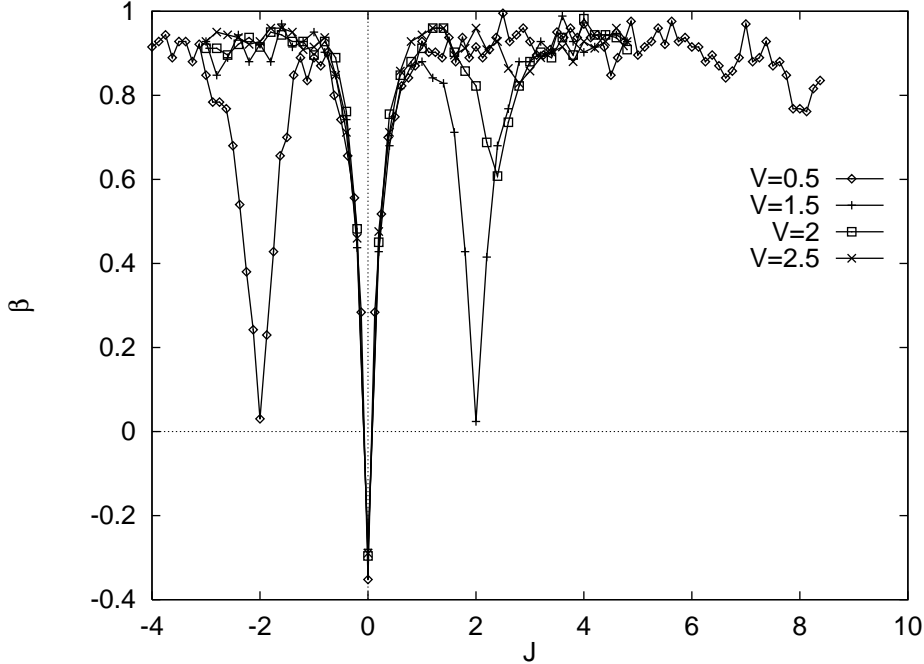} \par}

\caption{\label{FigHubbeta}Repulsion as a function of the coupling (see text).}
\end{figure*}

\subsection{The Chiral Quantum Potts Chain.}

We now turn to another quantum Hamiltonian : the quantum chiral Potts
chain Refs.\cite{vongehlenritten,howeskadanof}. The corresponding
transfer matrix has a \emph{higher genus} integrability Refs. \cite{auyangmaccoy,perkbaxterauyang}.
So it is natural to wonder if also the Gaussian ensembles provide
correct descriptions of the spectrum. The quantum Hamiltonian reads:\begin{equation}
\label{HPotts}
H=\sum _{j}\: \sum ^{N-1}_{n=1}\left[ \overline{\alpha _{n}}\left( X_{j}\right) ^{n}+\alpha _{n}\left( Z_{j}Z^{\dagger }_{j+1}\right) ^{n}\right] 
\end{equation}
where \( X_{j}=I\bigotimes \cdots \bigotimes X\bigotimes \cdots I \)
and \( Z_{j}=I\bigotimes \cdots \bigotimes Z\bigotimes \cdots I \)
operators X and Z are in position position \( j \), \( I \) is the
unit \( q\times q \) matrix, \( X_{ij}=\delta _{i,j+1}\;  \) mod(\( N) \)
and \( Z_{ij}=\delta _{i,j}\exp \left( 2\pi i(j-1)/N\right)  \).
Moreover Hamiltonian Eq. \ref{HPotts} is complex, at least for general
values of the parameters, and so one expects, \emph{a priori,} a GUE
statistic. An integrability condition has been found for this model
Ref.\cite{auyangmaccoy,perkbaxterauyang}. Restricting ourself to
values of the parameters which ensure that Hamiltonian Eq. \ref{HPotts}
is hermitian, we have performed a RMT analysis which allows us to
conclude that i) along the integrable variety the RDE is an adequate
description and ii) for generic point the GOE is the correct description.
Point ii) is quite surprising, since the GUE was expected. This means
that there exists a basis, independent of the parameters, in which
the Hamiltonian is \emph{real}. We have been able to find this basis
for sizes smaller than \( L=6 \). Note that this property implies
the existence of a unitary operator \( K \) which is extremely over-constrained
(see the introduction). From our numerical results, we conjecture
the existence of such a basis for any chain size \( L \).

\subsection{The three-dimensional Ising model.}

The three-dimensional Ising model is certainly one of the most challenging
model of lattice statistical physics. In particular the properties
of the critical point are debated. To clarify this question we have
performed a RMT analysis, see Ref.\cite{PhysREvESpin} . We start
with an anisotropic Ising model on a cubic lattice. In two directions
the couplings have the value \( K_{2} \) while, in the third, it
has the value \( K_{1} \). When \( K_{1}=K_{2} \) this is the usual
isotropic cubic Ising model, and when \( K_{1}=0 \) it reduces to
the isotropic two dimensional square lattice. We keep constant \( K_{2}=1 \)
and vary \( K_{1} \) in a range strarting from a small negative coupling
constant value to a value sufficiently large to be certainly larger
than the critical value which can be crudely evaluated by different
means. The results are summarized in Figure \ref{ising3d}.
\begin{figure*}
{\centering \includegraphics{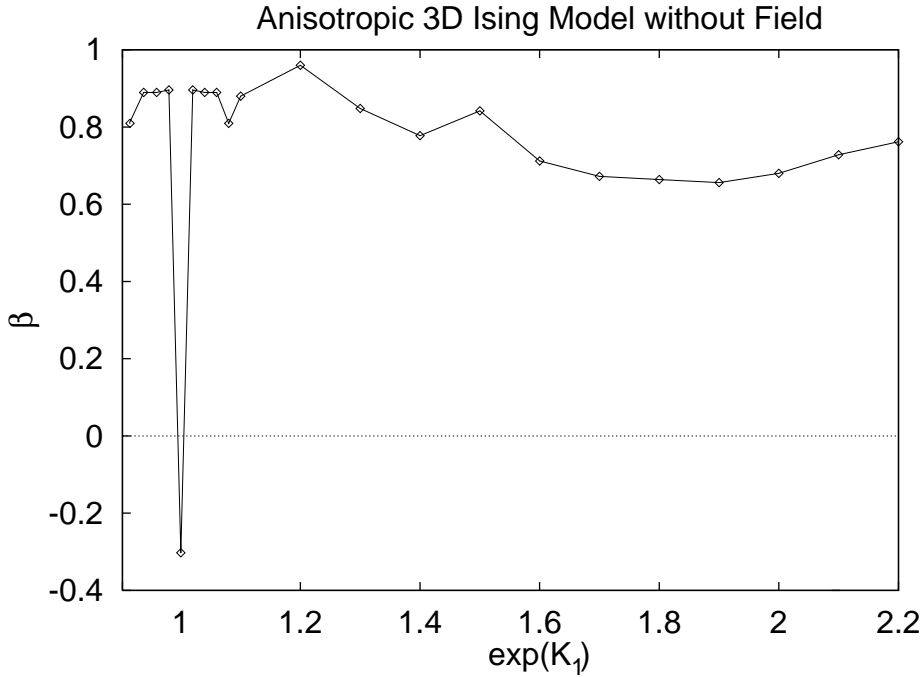} \par}

\caption{\label{ising3d}Three-dimensional Ising model}
\end{figure*}
It is clear, from this figure, that the critical point \emph{does
not} show any trace of a possible \emph{integrability-like property}%
\footnote{Let us recall that the reduction, in some scaling limit, of the \emph{critical}
three-dimensional model to some (integrable) conformal field theory,
thus yielding rational exponents, had been suggested by several authors.
}. This is in contrast with the results in the proximity of the two
dimensional case, where the absence of level repulsion is clearly
seen.

\subsection{The Ising model on the Kagomé lattice.}

To conclude this short note we would like to mention that we have
applied the RMT analysis to the Ising model on the Kagomé lattice.
F.Y. Wu, pointed out to us that it would be interesting to test the
critical point of the Kagomé lattice with the RMT analysis. The critical
manifold is not known, but F.Y. Wu conjectured some algebraic variety
for the anisotropic model see Ref. \cite{fanwu}. From this RMT analysis,
a value for the critical temperature of the isotropic model can be
deduced. This value is very close to the one obtained from Monte-Carlo
simulations. Fig. \ref{FigKag} (taken from Ref.\cite{kagomeJC})
presents the level spacing distribution for the conjectured critical
value \( K_{\textrm{W}} \), as well as for a generic value \( K=2 \).
It confirms a good agreement with the conjectured value, but, mainly,
it shows that the \emph{critical point is integrable}, in contrast
to the example of the three dimensional critical point.
\begin{figure}
{\centering \includegraphics{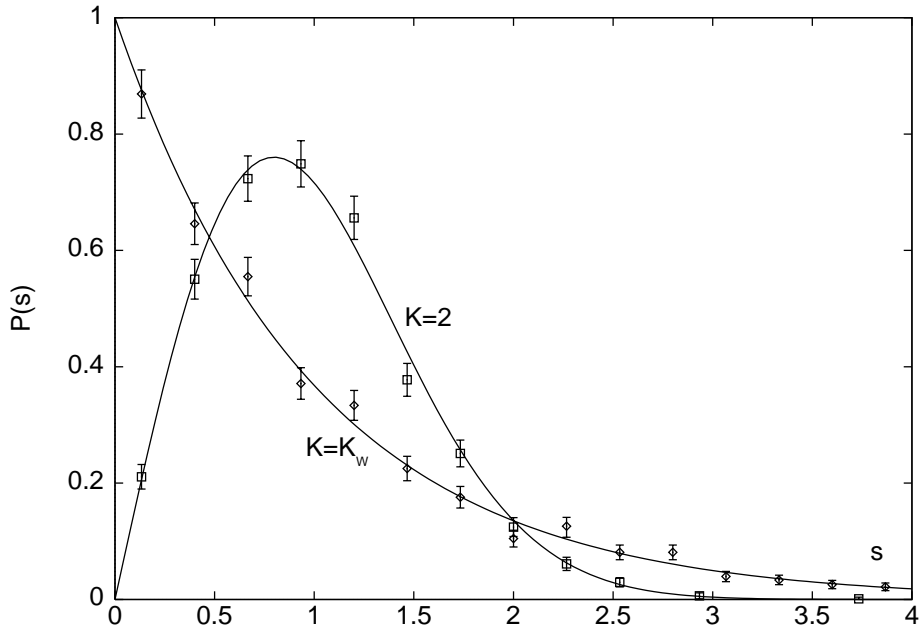} \par}

\caption{\label{FigKag}Ising model on a Kagomé lattice.}
\end{figure}

\section{Conclusions}

We have seen that RMT analysis could provide an alternative approach
to integrability and, to some extent, an alternative definition to
Bethe integrability or to Yang-Baxter integrability. It also gives
an operational way of testing integrability. We have found a time-reversal-like
unexpected symmetry in the Chiral Quantum Potts chain for which higher
genus integrability occurs, the spectrum being correctly described
by the RDE. Many other classical spin or vertex models, as well as
various quantum models, have also been investigated, all leading to
the same conclusions developed in this note. Furthermore we have shown
that the three-dimensional Ising model \emph{does} \emph{not} have
this property of independent eigenvalues for the spectrum of the transfer
matrices. This strongly suggests that this model will not be solved
without a genuinely new method, even at criticality.

\end{document}